\documentstyle[preprint,aps,epsfig]{revtex}
\begin{document}
\font\mybb=msbm10 at 12pt
\def\bb#1{\hbox{\mybb#1}}
\def\Z {\bb{Z}}
\def\R {\bb{R}}
\def\E {\bb{E}}
\def\unit{\hbox to 3.3pt{\hskip1.3pt \vrule height 7pt width .4pt \hskip.7pt
\vrule height 7.85pt width .4pt \kern-2.4pt
\hrulefill \kern-3pt
\raise 4pt\hbox{\char'40}}}
\def\II{{\unit}}
\def\cM {{\cal{M}}}
\def\half{{\textstyle {1 \over 2}}}

\def    \beq    {\begin{equation}} \def \eeq    {\end{equation}}
\def    \bea    {\begin{eqnarray}} \def \eea    {\end{eqnarray}}
\def    \lf     {\left (} \def  \rt     {\right )}
\def    \a      {\alpha} \def   \lm     {\lambda}
\def    \D      {\Delta} \def   \r      {\rho}
\def    \th     {\theta} \def   \rg     {\sqrt{g}} \def \Slash  {\, /
\! \! \! \!}  \def      \comma  {\; , \; \;} \def       \pl
{\partial} \def         \del    {\nabla}

\preprint{UG-00-01}

\title{A Noncommutative M--Theory Five--brane}
\author{E. Bergshoeff\footnote{e.bergshoeff@phys.rug.nl},
D. S. Berman\footnote{d.berman@phys.rug.nl},
J. P. van der Schaar\footnote{schaar@phys.rug.nl} and
P. Sundell\footnote{p.sundell@phys.rug.nl} }
\address{Institute for Theoretical Physics, University of Groningen, \\
Nijenborgh 4, 9747 AG Groningen, The Netherlands }

\maketitle
\begin{abstract}

We investigate, in a certain decoupling limit,
the effect of having a constant $C$--field
on the M--theory five--brane using an open membrane probe.
We define an open membrane metric for the
five--brane that remains non--degenerate in the limit.
The canonical quantisation of the open membrane boundary
leads to a noncommutative loop space which is a functional analogue of the
noncommutative geometry that occurs for D--branes.

\end{abstract}

\section{Introduction}

The M--theory five--brane is a most mysterious object
\cite{Dijkgraaf:1997cv,Howe:1997mx}. Several puzzles
remain concerning its effective action and the theory of coincident
five--branes, see e.g.~\cite{Bekaert:1999dp}.
Recently, there has been much interest in the noncommutative
geometry that arises on D--brane worldvolumes from including a constant 
Neveu
Schwarz two--form background potential
\cite{Connes:1998cr,Schwarz:1998qj,Seiberg:1999vs}. It is
natural to ask oneself whether similar deformed geometries play a
role in M--theory. It has been suggested that the five--brane of M--theory
plays the role of a D--brane in M--theory where, instead of an open string
ending on  the D--brane, we have an open membrane ending on the M5--brane
\cite{Strominger:1996ac,Townsend:1996af}. We note that the open
membrane ending on the five--brane can be reduced to the fundamental
string ending on the D$4$--brane and by dimensional reduction one may
therefore learn about the properties of D--branes by analysing the 
M5--brane, see for example \cite{Berman:1998va}.  The M--theory origin of 
the Neveu--Schwarz
two--form background potential is given by the three--form potential $C$ of
eleven--dimensional supergravity. In order to investigate the occurrence of
deformed geometries in M--theory one is therefore naturally led
to consider an M--theory open membrane/five--brane system
in the background of a constant $C$--field. The M5--brane in the background 
of a constant $C$--field has been investigated previously in the context of 
the AdS/CFT correspondence \cite{Maldacena:1999mh} and supersymmetric
n--cycles \cite{Lust:2000pq}.

It is known that, in order to derive the noncommutative properties of the 
D--brane, one may consider the interaction of open string end points with 
the Neveu--Schwarz two--form potential $B_{\rm NS}$ on the D--brane
and perform a canonical quantisation \cite{Chu:1999qz,Ardalan:1999av}.
Analogously, to determine the effect of including the background $C$--field 
on the usual five--brane geometry, we examine how an open membrane couples 
to an
eleven--dimensional supergravity background with background five--branes.
A preliminary analysis can be found in \cite{Chu:1999qz}.

In this paper we will not perform the full canonical quantisation programme 
to the open membrane. Inspired by the case of D--branes 
\cite{Connes:1998cr,Seiberg:1999vs}, we will consider a 
certain decoupling limit 
in which (1) the bulk modes of the membrane decouple; (2) the dynamics of 
the boundary string in the five--brane is governed by the Wess--Zumino term 
of the M2--brane action; and (3) the worldvolume theory of the M--theory 
five--brane is described
by a (2,0) supersymmetric field theory. Following this, we 
canonically quantise this 
Wess--Zumino boundary term only. The definition of the decoupling limit is 
more subtle than in the  case of strings \cite{Seiberg:1999vs} because: (i) 
there is  no coupling constant parameter $g_s$ in M-theory; (ii) the 
M5--brane couples to the dual of the $C$--field; and
(iii) 
the five--brane equations of motion involve a non--linear algebraic 
self--duality condition which must be taken into account \cite{Howe:1997mx}.

The canonical analysis of 
the boundary term leads to a noncommutative loop space on the five--brane 
worldvolume. This is a functional analogue of the noncommutative geometry 
that occurs for D--branes. Following this, we  give a suggestion for a 
loop space star product.

The decoupling limit mentioned above involves taking the field theory limit
on the five--brane in the presence of a background field strength. We 
conjecture
that the appropriate metric on the five--brane for describing this limit is 
not
the usual induced closed membrane metric but rather an {\it{open membrane 
metric}}. This metric is analogous to 
the
open string metric on a D--brane and indeed reduces to the latter upon 
double
dimensional  reduction. The open brane metrics are the natural metrics for 
describing the D--brane and the five--brane worldvolume theories in the
presence of a background field strength in that it is these metrics rather 
than the induced ones that give the linearised mass shell condition on the 
worldvolume which in turn defines the field theory limits on the branes.

The structure of the paper is as follows. In Section II we introduce the
open membrane/five--brane system. In Section III we discuss some properties 
of open brane metrics. Next, in Section IV, we give the decoupling limit and 
discuss its relation to the decoupling limit of  \cite{Seiberg:1999vs}. In 
Section V we derive the deformed five--brane geometry. First we apply the 
Dirac canonical quantisation programme to the decoupled boundary 
Wess--Zumino term and derive the Dirac brackets of the 
five--brane coordinates. We then connect to the results in the literature 
for the D4--brane, examine the special case of the infinite momentum frame
and finally give a star product for particular cases.  Finally, in Section 
VI we give  a discussion of our results and possible extensions to this 
work.

\section{The open membrane/five--brane system}

Our starting point is an open membrane in an eleven--dimensional supergravity
background with background five--branes. The membrane boundary is a string 
that
is constrained to lie within one of the five--branes which is separated from 
the stack of remaining five--branes. The reason that we include the 
additional stack of five--branes in the background is that it effectively 
stiffens the separated five--brane so that the membrane is not able to 
deform this five--brane and can be treated as a test membrane, as we shall 
discuss in more detail in Section IV.

The action for the open bosonic membrane is as follows (for the
kappa-symmetric version of this action, see \cite{Chu:1997iw}):

\beq
S = S_k + \int_{M^3} f_2^* C + \int_{\partial M^3} f_1^* b \ ,
\label{ac}
\eeq

where the kinetic term can be written in Polyakov form as

\beq
S_k={1\over 2(\ell_p)^2} \int_{M^3}d^3\sigma \sqrt{-\det{\gamma}}\left(
-\gamma^{\alpha\beta}\pl_{\alpha}X^M\pl_{\beta}X^N\hat{g}_{MN}+\ell_p^2\right)
\ .
\label{action}
\eeq

Here $\ell_p$ is the $D=11$ Planck's constant, $\hat{g}_{MN}$ is the $D=11$
spacetime metric and $\gamma_{\alpha\beta}$ is the auxiliary worldvolume
metric. The maps $f_2$ and $f_1$ denote the embedding of the membrane and 
its boundary into the spacetime and the five--brane, respectively. The 
worldvolume three--form $f_2^* C$ is the pull--back of the $D=11$ three--form
potential $C$ to the membrane worldvolume and, similarly, $f_1^* b$ is the 
pull--back of the five--brane two--form potential $b$ to the boundary of the
membrane. In terms of components, we write

\beq
(f_2^* C)_{\alpha\beta\gamma}= \pl_{\alpha} X^M \pl_{\beta} X^N \pl_{\gamma}
X^P
C_{MNP}\ ,\quad (f_1^* b)_{ij} = \partial_i X^\mu \partial_j X^\nu
b_{\mu\nu}\, ,
\eeq

where $M=0,1,\cdots,9,11$ are spacetime indices, $\mu=0, 1 \cdots ,5$ are  
five--brane worldvolume indices, $\alpha=0,1,2$ are membrane worldvolume 
indices and $i=0,1$ are indices on the boundary of the membrane.

Since the decoupling limit involves scaling Planck's constant $\ell_p$ we 
need to carefully assign dimensions to all quantities. We are using
units in which the spacetime metric $\hat{g}_{MN}$, the worldvolume metric 
$\gamma_{\alpha\beta}$, all differential forms and the membrane worldvolume 
parameters $\sigma^{\alpha}$ are dimensionless. The spacetime coordinates 
$X^M$ and the five--brane coordinates $X^{\mu}$ have dimension length. Note 
that the components $C_{MNP}$ and $b_{\mu\nu}$ have dimension (mass)$^3$ and 
(mass)$^2$, respectively, in order for $C$ and $b$ to be dimensionless.

The coupling of $b$ to the boundary of the membrane ensures that the open 
membrane action is invariant under the spacetime gauge transformations
$\delta C=d\Lambda$ provided that $\delta b= -f_5^* \Lambda$, where $f_5$ 
denotes the embedding of the five--brane into spacetime. The two--form $b$ 
satisfies the five--brane field equations. These are equivalent to a 
non--linear self--duality condition on the following gauge invariant three--form field
strength of $b$:

\beq
{\cal H}= db+f_5^* C
\ .\label{calh}
\eeq

Here the last term is the pull--back of the spacetime three--form potential 
to the five--brane:

\beq
(f_5^* C)_{\mu\nu\rho}= \pl_{\mu} x^M \pl_{\nu} x^N \pl_{\rho}
x^P
C_{MNP}\, ,
\eeq

where $x^M(X^\mu)$ are local embedding functions satisfying the
five--brane equations of motion. The non--linear self--duality condition on
the five--brane reads \cite{Howe:1997mx}

\beq
{\sqrt{-\det g}\over 6}\epsilon_{\mu\nu\rho\sigma\lambda\tau}
{\cal H}^{\sigma\lambda\tau}={1+K\over 2}(G^{-1})_{\mu}{}^{\lambda}
{\cal H}_{\nu\rho\lambda}\ ,
\label{nlsd}
\eeq

where $\epsilon^{012345}=1$ and the scalar $K$ and the tensor $G_{\mu\nu}$ are given by

\begin{eqnarray}
\label{k}
K &=&\sqrt{1+{\ell_p^6 \over 24}{\cal{H}}^2}\, ,\\
&&\cr
G_{\mu\nu} &=& {1+K\over 2K}\left(g_{\mu\nu}+{\ell_p^6\over 4}
{\cal H}^2_{\mu\nu}\right)\ .
\label{om}
\end{eqnarray}

We shall argue in the next section that the tensor $G_{\mu\nu}$ is the 
metric on the five--brane seen by an open membrane in the presence of a 
background three--form field strength ${\cal H}$. 
It is also understood that in 
the above three equations the indices are contracted with the induced 
five--brane metric:

\beq
g_{\mu\nu}=\pl_{\mu}x^M\pl_{\nu}x^N\hat{g}_{MN}\ .
\label{gmunu}
\eeq

We may parameterise the $D=11$ spacetime by local 
coordinates $X^M=(X^{\mu},Y^m)$, where $m=6,7,8,9,11$ refer to the 
directions perpendicular to the five--brane, such that the Dirichlet 
condition on the membrane embedding fields becomes 
$Y^m(\sigma^{\alpha})|_{\pl M^3}=0$.
The remaining parallel embedding 
fields $X^{\mu}(\sigma^{\alpha})$ obey a mixed Dirichlet and Neumann 
boundary condition. By varying the action (\ref{ac}), using $\delta 
Y^m|_{\pl M^3}=0$ together with the 
definitions (\ref{calh}) and (\ref{gmunu}) one finds that \cite{Chu:1999qz} 

\beq
\label {bc}
\left.\left[{1\over \ell_p^2} \sqrt{-\det{\gamma}}\, n^{\alpha}
\pl_{\alpha}X^{M}{\hat g}_{\mu M}
+\epsilon^{\alpha\beta\gamma} n_{\alpha}
{\cal H}_{\mu\nu\rho} \pl_{\beta} X^{\nu} \pl_{\gamma} X^{\rho}
\right] \right|_{\pl M^3}=  0\ ,
\eeq

where $n_{\alpha}$ is the normal vector at the boundary. The first term is 
non--local from the point of view of the boundary string. As we shall see, 
this term drops out in the decoupling limit leaving a local field equation 
for the embedding fields $X^{\mu}(\sigma^i)$ of a closed string in six 
dimensions. The closed string boundary conditions are

\beq
X^{\mu}(\tau,\sigma+2\pi)=X^{\mu}(\tau,\sigma)\ ,
\label{csbc}
\eeq

where $\tau\equiv\sigma^0$ is the time coordinate and $\sigma\equiv\sigma^1$ 
is the spatial coordinate on the worldsheet.

We shall consider backgrounds where ${\cal H}_{\mu\nu\rho}$ is constant. 
This is only consistent with (\ref{calh}) provided we require that the 
pull--back of the spacetime four--form field strength $F=dC$ to the 
five--brane vanishes, i.e.~$f_5^\star F = 0$. After some manipulations 
(given in appendix A) we end up with the following action:

\beq
\label{form}
S= S_k+ \int_{M^3}f_2^*\tilde{C}+ {1\over 3}\int_{\partial M^3}  d^2\sigma\,
{\cal H}_{\mu \nu \rho}
X^{\mu} {\dot X}^{\nu}  X^{\prime\rho} \ ,
\eeq

where $\tilde{C}$ is the part of the three--form potential which is 
perpendicular to the five--brane, i.e.~$f_5^\star \tilde{C} = 0$, and the 
dot and prime indicates differentiation with respect to $\tau$ and $\sigma$, 
respectively.

It will be useful to introduce a specific parametrization
of the solutions of the self--duality condition (\ref{nlsd})
as follows\footnote{This parameterisation has been derived 
independently by \cite{MC}.}  (see Appendix B):

\beq
{\cal H}_{\mu\nu\rho}=  {h\over
\sqrt{1+\ell_p^6 h^2}} \epsilon_{\alpha\beta\gamma}
v^{\alpha}_{\mu} v^{\beta}_{\nu} v^{\gamma}_{\rho}
+h\, \epsilon_{abc}
u^{a}_{\mu} u^{b}_{\nu} u^{c}_{\rho}\ ,
\label{hsol}
\eeq
\beq
G_{\mu\nu}= {\left(1+\sqrt{1+h^2\ell_p^6}\right)^2\over 4}\left(
{1\over 1+h^2\ell_p^6}\eta_{\alpha\beta}v^{\alpha}_{\mu}v^{\beta}_{\nu}
+\delta_{ab}u^{a}_{\mu}u^{b}_{\nu}\right)\ .
\label{openm} \eeq

Here $h$ is a real field of dimension (mass)$^3$ and $(v_\mu^\alpha, 
u_{\mu}^{a})$, $\alpha=0,1,2$, $a=3,4,5$, are sechsbein fields in the 
nine--dimensional coset $SO(5,1)/SO(2,1)\times SO(3)$ satisfying

\beq
g^{\mu\nu}v_{\mu}^{\alpha}v_{\nu}^{\beta}=\eta^{\alpha\beta}\ ,\quad
g^{\mu\nu}u_{\mu}^{a}v_{\nu}^{\beta}=0\ ,\quad
g^{\mu\nu}u_{\mu}^{a}u_{\nu}^{b} = \delta^{ab}\ , \nonumber
\eeq
\beq
g_{\mu\nu} = \eta_{\alpha\beta}v_\mu^\alpha v_\nu^\beta
+ \delta_{ab} u_\mu^{a}u_\nu^{b}\, .
\label{sol}
\eeq

\section{Open Brane Metrics}

The natural metric to examine the worldvolume of a D--brane with a non--trivial
background Born--Infeld field strength ${\cal F}_{\mu\nu}$ is the so called
open string metric. For a D$p$--brane this metric is defined as follows:

\beq
G_{\mu\nu} = g_{\mu\nu} - {\alpha}'^2{\cal{F}}_{\mu\rho} g^{\rho\lambda}
{\cal{F}}_{\lambda\nu}\, ,\hskip 1.5truecm  \mu = 0,1,\cdots ,p\, ,
\label{os}
\eeq

where $g_{\mu\nu}$ is the induced metric on the brane. The open string 
metric defines the mass shell condition for the open string modes 
propagating in the D--brane\cite{Seiberg:1999vs}. The expression (\ref{os}) for
the open string metric can also be read off from the low energy effective 
Dirac--Born--Infeld action which yields massless field equations with 
D'Alembertians $(G^{-1})^{\mu\nu}\nabla_{\mu}\nabla_{\nu}$ and Dirac 
operators $(G^{-1})^{\mu\nu}\Gamma_{\mu}\nabla_{\nu}$ (where $\nabla_{\mu}$ 
contains the Christoffel symbol of $g_{\mu\nu}$). The decoupling limit
on a D$p$--brane as defined in \cite{Seiberg:1999vs}
is non--degenerate in the sense that it yields a  
$(p+1)$--dimensional, massless field theory on the brane if the leading 
components of the open string metric form a rank $p+1$ matrix.

In analogy with how branes appear as solutions in supergravity, the open 
string is visible as a supersymmetric worldvolume soliton solution to the 
massless D--brane field equations. It is instructive to examine the geometry
of these so called BIon solutions \cite{Callan:1998kz} using the open string 
metric. A charged, static BIon, corresponding to the string ending on a 
D$p$--brane, excites one transverse scalar X and the time component of the 
Born-Infeld vector field $A_{\mu}$. The solution is given by

\beq
(\alpha')^{{1\over 2}}A_0= (\alpha')^{-{1\over 2}}X = H\ ,
\eeq

where $H$ is a harmonic function on the transverse worldvolume space $\E^p$ 
given by:

\beq
H=1+ {Q \over r^{p-2}}\, , \qquad p>2\ .
\eeq

After substituting this solution into the open string metric (\ref{os}) we 
find the following two--block 0--brane line element:

\beq
ds^2(G)= -{dt^2\over 1+\alpha'(\pl H)^2} + (dy^m)^2\ ,\hskip 1.5truecm m=1,
\cdots p\, ,
\eeq

where $(\pl H)^2=\delta^{mn}\pl_m H \pl_n H$ and $(dy^m)^2$ is the line 
element on $\E^p$. Taking the limit of large $Q$ and carrying out the 
coordinate transformation $r^{2(p-2)}= \alpha' Q^2/u^2$ we find the 
following metric:

\beq
ds^2(G) ={r^2 \over (p-2)^2 } \lf { -dt^2 +du^2 \over u^2} +(p-2)^2
d\Omega^2_{p-1}\rt \ .
\eeq

This metric is conformally $AdS_2 \times S^{p-1}$. The ordinary induced 
metric $ds^2(g)$ is 
of three--block form with an additional line element proportional to $dr^2$ 
and therefore not conformally $AdS_2\times S^{p-1}$ in the near horizon  
region.

The natural properties of the open string metric (\ref{os}) in examining the 
open string spectrum, motivates the introduction of an analogous
metric for studying an open membrane probing a five--brane in the
presence of a background ${\cal H}$ field. We propose that the conformal 
class of this so called open membrane metric is given by the tensor
$G_{\mu\nu}$ given in (\ref{om}). This
metric has been shown to dimensionally reduce to the open string metric on 
the D4--brane up to a scale factor \cite{Howe:1997mx}.
Moreover, the massless field equations on the five--brane have kinetic terms 
involving the box operator $(G^{-1})^{\mu\nu}\nabla_{\mu} \nabla_{\nu}$ and 
the Dirac operator $(G^{-1})^{\mu\nu}\Gamma_{\mu}\nabla_{\nu}$ 
\cite{Howe:1997mx}. Further arguments in favour of the proposed open 
membrane metric are obtained by considering other excitations than the 
massless tensor multiplet, like e.g.~the self--dual string soliton 
corresponding to the open membrane ending on the 
five--brane\cite{Howe:1998ue}\footnote{Only in this section we use the index 
$m$ to indicate the five--brane worldvolume directions transverse to the 
self--dual string. In all other sections the index $m$ indicates the target 
space directions transverse to the five--brane.}:

\beq
4\, \ell_p^2\, b_{01} = \ell_p^{-1} X =  H\, , \qquad
{\cal H}_{mnp}= {\ell_p^{-2}\over 4} \epsilon_{mnpq}
\pl_q H\, ,\hskip 1truecm m=2,3,4,5\, ,
\eeq

where $H$ is a harmonic function on the transverse worldvolume space $\E^4$ 
given by:

\beq
H=1+ {Q \over {r^2}}\, .
\eeq

Substituting this solution into (\ref{om}) the line element of the open 
membrane metric has the structure of a two--block 1--brane solution:

\beq
ds^2(G)={1\over 4}\left(1+\sqrt{1+\ell_p^2(\pl H)^2}\right)^2
\left({-dt^2+dx^2\over 1+\ell_p^2(\pl H)^2}+ (dy^m)^2\right)\ ,
\eeq

whereas the induced metric has an additional line element proportional to 
$dr^2$. In the large $Q$ limit the open membrane metric, upon making the 
coordinate transformation $r^4 = \ell_p^2 Q^2/u^2$, becomes:

\beq
ds^2(G)= {u^2\over 4} \left (  { -dt^2 + d\sigma^2 +
du^2  \over u^2} + 4 d\Omega^2_3 \right )\, ,
\eeq

which we identify as $AdS_3 \times S^3$ with a conformal factor.

\section{The decoupling limit}

We will now consider a limit of the open membrane/five--brane system with 
the following three properties. Firstly, the bulk modes of the membrane can 
be neglected. Secondly the boundary string that lives in the five--brane is 
governed solely by the Wess--Zumino term. Thirdly, the field theory limit on 
the five--brane is taken such that the physics on the five--brane is 
described by the self--dual tensor field theory. In order to demonstrate the 
limiting procedure we require, it is instructive to first examine the the 
limit taken by Seiberg and Witten in deriving the noncommutative geometry of 
a D--brane from the open string.

\subsection*{The open string case}

We study the decoupling limit at small string coupling $g_s$. The effective 
tensions $\tau$ of the string and the D$p$--brane behave like $\tau_{\rm F1} 
\sim 1,\, \tau_{\rm Dp} \sim 1/g_s$. Therefore, for small $g_s$, the string is 
much lighter than the D$p$--brane and can be treated as a test string probing
the D$p$--brane. Furthermore, the effective gravitational couplings 
$G_N\tau$ (Newton's constant times tension) behave like $G_N\tau_{\rm F1} 
\sim g_s^2, G_N\tau_{\rm Dp} \sim g_s$ and therefore we can assume that the 
spacetime background is approximately flat. The open string action reads

\beq
S={1\over 2\alpha'} \int_{M^2} d^2\sigma\,  \bar \pl X^{M}
\pl X^{N} \eta_{MN}
+{1\over 2}\int_{\pl M^2}d\tau {\cal F}_{\mu\nu} X^{\mu}\dot{X}^{\nu}\ ,
\label{stringa}
\eeq

where ${\cal F}_{\mu\nu}$ is the background field strength on the 
D$p$--brane. We assume that the only non--vanishing components of ${\cal F}$ 
are ${\cal F}_{r^{\prime} s^{\prime}}$, where we have decomposed the
worldvolume index $\mu$ as    $\mu = (r,r^\prime)$ with 
$r = 0,1,\cdots ,p- {\rm rank}\, {\cal F}$ and
$r^\prime = p+1-{\rm rank}\, {\cal F} ,\cdots,p$.

The string theory has closed string modes in the bulk, i.e.~away from the 
D--brane. Their mass shell condition is governed by the closed
string metric $\eta_{MN}$:

\beq
\eta^{MN}p_M p_N=-{4\over \alpha^\prime}\bigl ( N_{\rm L} -1\bigr ) =
-{4\over \alpha^\prime}\bigl ( N_{\rm R} -1\bigr )\, .
\eeq

Here $N_{\rm L,R}$ indicates the oscillator number of the left and right 
movers and $p_M$ is the bulk momentum of the closed string state. There are 
also open string modes propagating in the D--brane. An open string state 
with worldvolume momentum $k_{\mu}$ obeys the mass shell condition

\beq
(G^{-1})^{\mu\nu}k_{\mu}k_{\nu}=-{1\over \alpha^\prime} \bigl(
N^{\rm open}-1\bigr )\ ,
\eeq

where $N^{\rm open}$ is the open string oscillator number and the tensor 
$G_{\mu\nu}$ is the open string metric defined in (\ref{os}). From the 
conditions on ${\cal F}_{\mu\nu}$, it follows that $G_{\mu\nu}$ is given by

\beq
G_{\mu\nu}=\left\{\begin{array}{ll}
\eta_{\mu\nu} &\ \ \mbox{for}\ \ \mu,\nu = r,s\\
\eta_{\mu\nu}-\alpha^{\prime 2}{\cal F}_{\mu\rho}\eta^{\rho\sigma}
{\cal F}_{\sigma\nu}& \ \ \mbox{for}\ \
\mu,\nu= r^\prime,s^\prime
\end{array}\right.
\eeq

In order to describe the limit we first decompose the target spacetime index 
$M$ as $M = (r,r^\prime,m)$ and split the spacetime metric $\eta_{MN}$ into 
parts that are parallel and perpendicular to ${\cal{F}}$ and the D--brane as
follows: $\eta_{MN} \rightarrow \eta_{rs} \oplus \eta_{r^\prime s^\prime} 
\oplus \eta_{mn}$, where 
$\eta_{rs}$ is perpendicular to ${\cal F}$ and
parallel to the D--brane, $\eta_{r^\prime s^\prime}$
is parallel to ${\cal F}$,
  and $\eta_{mn}$ is perpendicular to the D--brane.

The limit of Seiberg and Witten is obtained by taking $\epsilon\rightarrow 
0$ such that\footnote{In the equation below it is understood that the 
$\eta_{r^\prime s^\prime }$ and $\alpha^\prime$ occurring at the 
right--hand--side are 
$\epsilon$--independent.}:

\beq
\eta_{r^\prime s^\prime }  \sim  \epsilon\, \eta_{r^\prime s^\prime} 
\ , \quad \alpha'
\sim \epsilon^{{1\over 2}}\alpha^\prime  \  , \label{swl}
\eeq

while keeping all other quantities fixed. The open string action scales as 
follows:

\begin{eqnarray}
S &\sim& {1\over 2\epsilon^{1/2}\alpha^\prime} \int_{M^2} d^2 \sigma ~
\bar \pl X^{m} \pl X^{n} \eta_{mn} +
{1\over 2\epsilon^{1/2}\alpha^\prime} \int_{M^2} d^2 \sigma ~
\bar \pl X^{r} \pl X^{s} \eta_{r s}\nonumber\\
&&\\
&&+ {\epsilon^{1/2}\over 2\alpha^\prime}
\int_{\Sigma} d^2 \sigma ~ \bar \pl X^{r^\prime }\pl X^{s^\prime } 
\eta_{r^\prime s^\prime }
+ {1\over 2}\int_{\pl M^2}d\tau {\cal F}_{r^\prime s^\prime } X^{r^\prime}
\dot{X}^{s^\prime }\nonumber
\end{eqnarray}

and the various string mass shell conditions become

\bea
{\rm bulk}\quad :\quad &&\quad\left\{ \begin{array}{l}
\eta^{mn}p_mp_n\ ,\quad  \eta^{r s}p_{r}p_{s}
\sim
-{4\over \epsilon^{1/2} \alpha^\prime }(N-1)\, ,\qquad
N=N_{\rm L} = N_{\rm R}\, ,\\
\eta^{r^\prime s^\prime }p_{r^\prime}p_{s^\prime}
 \sim - {4\epsilon^{1/2}\over \alpha^\prime}(N-1)\, ,
\end{array}\right. \nonumber\\
{\rm D-brane}\quad :\quad && \quad(G^{-1})^{\mu\nu}k_{\mu}k_{\nu} \sim -{1\over
\epsilon^{1/2}\alpha^\prime} (N^{\rm open}-1) \ ,\nonumber
\label{osm}
\eea

where the open string metric is finite in this limit and is given by the 
maximal rank matrix

\beq
G_{\mu\nu}=\left\{ \begin{array}{ll}
\eta_{\mu\nu} & \mbox{for}\ \ \ \ \mu,\nu = r, s\\
-\alpha'^2{\cal F}_{\mu\rho}\eta^{\rho\sigma}{\cal F}_{\sigma\nu} &
\mbox{for}\ \ \ \ \mu,\nu=r^\prime,s^\prime
\end{array} \right. 
\eeq

The limit (\ref{swl}) has the following three crucial properties that we 
wish to emulate for the open membrane/five--brane system:

\begin{enumerate}

\item[i)] The closed string bulk modes perpendicular to ${\cal F}$ with 
momentum $p_M = (p_r,0, 0)$ or $p_M = (0,0,p_m)$ are frozen out\footnote{
The massless, perpendicular bulk 
modes remain but their effective bulk coupling constant goes to zero.}
thereby isolating the dynamics of the D--brane theory from the bulk.

\item[ii)] The closed string bulk modes with momentum 
$p_M=(0,p_{r^\prime},0)$ give vanishing contribution to the action.

\item[iii)] On the D--brane all massive open string modes are frozen out and 
the decoupled, massless field theory on the brane is non--degenerate in the 
sense that it is $(p+1)$--dimensional and has finite effective coupling.

\end{enumerate}

The noncommutative nature of the D--brane arises from quantising the
remaining Wess--Zumino term.

\subsection*{The open membrane case\label{sec:dl}}

We now would prefer to proceed by analogy to the open membrane/five--brane 
system. However, the analysis of the decoupling limit for this system 
requires a slight modification, due to three circumstances: the absence of 
an analog of the string coupling constant; the fact that the open membrane 
probe couples to the three--form potential which is dual to the potential of 
the background five--brane; and the non--linear self--duality condition on 
the field strength in the five--brane worldvolume.

Thus, in M-theory, there is no sense in which the tension of an isolated 
five--brane can be said to be much larger than the tension of an isolated 
membrane. Neither can we assume an approximately flat spacetime background around a
background five--brane. To prevent the membrane from
deforming the
background geometry, we therefore consider a $D=11$ background consisting of 
a large stack of parallel five--branes, given by the solution

\beq
ds^2(\hat{g})= H^{-1/3}(dx^\mu)^2 + H^{2/3} (dy^m)^2\ ,\quad
H=1+{N_5\ell_p^3\over r^3}\ ,\quad
F=N_5\epsilon_4\ , \label{metric}
\eeq

where $\mu = 0,1,\cdots ,5;\, m = 6,7,8,9,11,\, N_5$ is the number of 
stacked five--branes and $\epsilon_4$ is the volume form on the transverse 
$S^4$. We let the open membrane end on one of these five--branes removed 
from the stack and placed at radius $r_0$. If $N_5>>1$ and $r_0$ is small, 
then the interactions between the stack and the separated five--brane 
effectively stiffens the latter so that the membrane can probe it without 
deforming it\footnote{ 
By analysing the Nambu--Goto action for the five--brane scalars, 
it can be shown that for large $N_5$ the characteristic length scale of 
five--brane deformations with large gradients is $N_5^{-{1 \over 3}} \ell_p$.}.

Under these conditions the induced metric on the five--brane (\ref{gmunu}) 
is given by:

\beq
g_{\mu\nu}=H^{-1/3}(r_0)\eta_{\mu\nu}\ .
\label{gmn}
\eeq

Moreover, from (\ref{metric}) it follows that the $D=11$ background four--form field
strength satisfies $f_5^*F=0$. From the discussion in Section II it
follows that we may consider an open membrane action given by

\bea
\label{oma}
S&=&{1\over 2(\ell_p)^2} \int_{M^3}d^3\sigma\sqrt{-\det{\gamma}}\left(
-H^{-1/3}\gamma^{\alpha\beta}\pl_{\alpha}X^{\mu}\pl_{\beta}X^{\nu}\eta_{\mu\nu}
-H^{2/3}\gamma^{\alpha\beta}\pl_{\alpha}Y^m\pl_{\beta}Y^n\delta_{mn}
+\ell_p^2\right) \nonumber\\
&&+N_5\int_{M^3} f_2^\star \tilde{C}+{1\over 3}\int_{\pl M^3} d^2\sigma\,
{\cal  H}_{\mu\nu\rho}
X^{\mu} {\dot X}^{\nu} X^{\prime\rho}\ ,
\eea

where the $D=11$ background three--form potential $\tilde{C}$ obeys $d\tilde{C}=\epsilon_4$
and $f_5^*\tilde{C}=0$ and the background three--form field strength
${\cal H}_{\mu\nu\rho}$ on the five--brane is constant.

Unlike the case of the string ending on a D--brane, where the Neveu--Schwarz two--form
potential can be tuned independently of the Ramond--Ramond potentials, the membrane
couples to the dual potential $N_5\tilde{C}$ of the five--brane six--form 
potential, which becomes large in the limit of large $N_5$. However,
$\tilde{C}$ affects only the perpendicular closed membrane bulk modes and does
not prevent these modes from decoupling in the limit from the dynamics on the
five--brane. We will therefore, from now on, drop the term in the open
membrane action (\ref{oma}) containing $\tilde{C}$.

Another difference from the string case is that we cannot with impunity take 
a scaling limit for ${\cal H}$, since the field equations for the 
five--brane impose the non--linear algebraic constraints (\ref{nlsd}). In 
order for our limit to be guaranteed to obey (\ref{nlsd}) we will use the 
explicit solution for ${\cal H}$ given by (\ref{hsol}) and (\ref{sol}).

We propose the decoupling limit obtained by taking $\epsilon\rightarrow 0$ 
such that

\bea
\ell_p&\sim& \epsilon\, \ell_p\ ,\\
{N_5\over r_0^3}&\sim& \epsilon^{-3\delta}{N_5\over r_0^3}\ ,\label{nr}\\
h&\sim& \epsilon^{-\lambda}h\ .
\label{lambda}
\eea

For simplicity we shall assume that $\delta> 1$\footnote{
One can also consider the case  $0<\delta\leq 1$. 
This does not lead to any 
qualitative changes in the final result.}
such  that we may drop the 
$1$ from the harmonic function in the metric (\ref{metric}) \footnote{We
point out that the limit considered here differs from the Maldacena limit
since energies are not kept fixed in the near horizon region. }. It then
follows from (\ref{gmn}) that the induced five--brane metric  and the
sechsbein fields in (\ref{sol}) scales as

\bea
g_{\mu\nu} \sim \epsilon^{\delta - 1}\, g_{\mu\nu}\ ,\quad u^a_{\mu}  \sim
\epsilon^{{1\over 2}(\delta-1)}u^a_{\mu} \ , \quad v^{\alpha}_{\mu}  \sim
\epsilon^{{1\over 2}(\delta-1)}v^{\alpha}_{\mu}\ .
\label{gmns}
\eea

Furthermore, we assume that $\lambda \le 3$. This implies that  $h\ell_p^3$ 
remains finite which enables us to keep the three--form field strength
and the open membrane metric non--degenerate in the limit. Thus
we find that the open membrane action (\ref{oma}) scales as

\bea
S&\sim&\epsilon^{-\Delta}\left[{1\over 2\ell_p^2}
\int_{M^3}d^3\sigma\sqrt{-\gamma}\left(
-\epsilon^{\Delta+\delta-3} H^{-{1\over 3}}\gamma^{\alpha\beta}
\pl_{\alpha}X^{\mu} \pl_{\beta}X^{\nu} \eta_{\mu\nu}
-\epsilon^{\Delta- 2\delta}H^{{2\over 3}}
\gamma^{\alpha\beta} \pl_{\alpha}Y^m \pl_{\beta}Y^n \delta_{mn}+
\epsilon^{\Delta}\ell_p^2\right) \right.\nonumber\\
&&\left.+ {1\over 3} \int_{\pl M^3}d^2\sigma\,
{\cal H}_{\mu\nu\rho}
X^{\mu} {\dot X}^{\nu} X^{\prime\rho}\right]\ ,
\label{es}
\eea

where we have defined

\beq
\Delta=\lambda-{3\over 2}(\delta-1)\ .
\eeq

We now make our requirements, as for the string case:

\begin{enumerate}

\item[i)] The perpendicular bulk modes $Y^m$ are frozen out provided 
$\Delta-2\delta<0$.

\item[ii)] The action for the parallel bulk modes vanishes if
$\Delta+\delta-3>0$. (This amounts to the
vanishing of the first term in (\ref{bc}) such that  (\ref{bc}) turns into a
local field equation on the string worldsheet.)

\item[iii)] We require a non--degenerate field theory limit on the 
five--brane. As discussed in Section III, we conjecture that the relevant 
metric is the open membrane metric (\ref{om}). Thus we require the distances 
$ds^2(G)$ measured in the open membrane metric to scale more slowly than 
$\ell_p^2$ and the leading part of the open membrane metric to be 
non--degenerate in the limit\footnote{This differs from the 
limit proposed in \cite{Seiberg:1999vs} where the rank of the 
open membrane metric is reduced from six to five in the limit.}:

\beq
G_{\mu\nu}\sim \epsilon^{\beta}\left(\bar{G}_{\mu\nu}+{\cal
O}(\epsilon)\right)\ ,
\label{glimit}
\eeq

where

\beq
\beta<2\ ,\qquad \det \bar{G}_{\mu\nu}\neq 0\ .
\label{omc}
\eeq

From (\ref{openm}) it follows that the second condition in (\ref{omc})
requires $h\ell_p^3$ to remain finite in the limit which implies that 
$\lambda \le 3$. The first condition in (\ref{omc}) amounts to $\delta<3$.

\end{enumerate}

Combining our assumption that $\delta >1$ with the conditions obtained under 
(i)-(iii) we find the following restrictions on our parameters:

\beq
\Delta+{3\over 2}(\delta-1)\leq 3< \Delta+\delta\ ,\quad \Delta<2\delta\ ,
\quad  1<\delta<3\ .  \label{limsum}
\eeq

These conditions are solved by $(\Delta,\delta)$ in a finite size region. 
For instance, $\delta={5\over 3}$, $\lambda=3$ and $\Delta=2$ leads to a 
decoupled five--brane theory in a background with a non--linearly self--dual 
field strength, while $\delta={4\over 3}$, $\lambda=2$ and $\Delta=2$ yields 
a linearly self--dual field strength. 

A noteworthy feature is that 
(\ref{limsum}) implies $\Delta>0$, such that there is 
necessarily an overall scaling of
the action in (\ref{es}). Such a scaling was not required in the
string case. A crucial difference between the string and the membrane
is that only the string action (\ref{stringa}) has a 
microscopic interpretation. On the other hand, the membrane action
should be seen as an effective action.
One interpretation of the scaling  (\ref{es}) with $\Delta > 0$ 
is that actually we are taking a semiclassical limit. 

Summarizing, in order to understand the geometry of the five--brane 
worldvolume we are led to study the quantisation of the following action:

\beq
\label{wza}
S= {1\over 3}\int_{\partial M^3} d^2\sigma\,
{\cal H}_{\mu \nu \rho} X^{\mu}
{\dot X}^{\nu} X^{\prime\rho}\ , \label{action0}
\eeq

with ${\cal H}$ constant.

\subsection*{A comparison with the limits for a lifted D--brane}

Before proceeding with the canonical analysis in the next section, we wish 
to compare the decoupling limit we propose for the M-theory five--brane with 
the one discussed in \cite{Seiberg:1999vs}. By applying the usual relations 
between string theory and M-theory

\beq
g_s=(R/\ell_p)^{3/2}\ , \quad \alpha' = { \ell_p^3 \over R}\, ,\quad
{\cal F}_{\mu\nu}= R\, {\cal H}_{\mu\nu5}\, ,\quad  \mu,\nu = 0,1,2,3,4\, ,
\eeq

one may lift the limits taken for the D4--brane to the case of the 
M5--brane. We assume that the rank  of ${\cal F}$ is $4$. This motivates the 
following limit for the M--theory five--brane on ${\bf{R}}^6$ 
\cite{Seiberg:1999vs}:

\bea
g_{IJ} &\sim & \epsilon\, g_{IJ}\, , \quad g_{55} \sim \epsilon^2 g_{55}\, ,
\quad g_{00}
\sim -1\, ,\quad  \ell_p \sim \epsilon^{1/2}\ell_p \, , \quad
I = 1,2,3,4\, , \nonumber\\
{\cal H}_{012} &\sim & \epsilon^{-1}\sqrt{1+\epsilon}\, h_0
\, , \quad {\cal H}_{034}=-\epsilon^{-1}h\, ,
\quad
{\cal H}_{125} \sim -h_0\, , \quad {\cal H}_{345} \sim  \sqrt{1+ \epsilon}\,
h \, ,
\eea

with $h_0^2 - h^2 + \ell_p^6 h_0^2 h^2 =0$ and $h_0<0$.

The presented constant three--form solution is  a special case of 
(\ref{hsol}) where a diagonalised ${\cal H}_{012} \oplus {\cal H}_{345}$ 
split background has been boosted in an $\epsilon$ dependent way in the 
$0-5$ direction which turns on ${\cal H}_{034}$ and ${\cal H}_{125}$ 
components. This is required in order for the reduction along the fifth 
direction to give a rank $4$ constant two--form solution on the D4--brane. 
Equivalently, we can reduce along a skew spacelike direction in the $0-5$ 
plane instead of reducing along the fifth direction to obtain a rank $4$ 
solution on the D4--brane.

By construction, the dimensional reduction of this solution and limit will 
give the appropriate limit for the D4--brane. However, under this limit the 
open membrane metric (\ref{openm}) behaves like $G_{55} \ell_p^{-2}  
\rightarrow  0$. This means that our condition (iii) for the field theory 
limit (involving the open membrane metric) on the five--brane to be valid is 
not satisfied. Obviously, this is  not a problem for the compactified theory 
where the excitations in the fifth direction are suppressed.

\section{Canonical Analysis}

In this section we will canonically quantise the action (\ref{wza}) with 
constant field strength ${\cal H}_{\mu\nu\rho}$. In the first part
of this section we will assume that the field strength 
 can be diagonalised as follows
(see Appendix B for details) \cite{Seiberg:1999vs}:

\beq
{\cal H}_{012} = -{ h \over \sqrt{1+ \ell_p^6 h^2} }\ , \qquad
{\cal H}_{345} = h\, .
\label{33}
\eeq

where the dimensionless tensor multiplet ``coupling'' $h\ell_p^3$ is 
non--vanishing provided the decoupling limit (\ref{lambda}) has been taken 
with $\lambda=3$. For $\lambda<3$ the limit results in a linear tensor 
multiplet. At the end of this section we discuss the case in which the 
field strength
cannot be brought into the above form (see Appendix C for details).

In the parameterisation (\ref{33}) the action (\ref{action0}) splits into 
two independent Lagrangians for the two sets of coordinates $X^{0,1,2}$ and
$X^{3,4,5}$:

\beq
S= {h\over 3\sqrt{1+ \ell_p^6 h^2}} \int_{\partial M^3} d^2\sigma\,
\epsilon_{\alpha\beta\gamma} X^{\alpha}
{\dot X}^{\beta}  X^{\prime\, \gamma} +{h \over 3} \int_{\partial M^3}
d^2\sigma\,
\epsilon_{abc} X^{a}
{\dot X}^{b}  X^{\prime\, c}  \ , \label{action1}
\eeq

where $\alpha=0,1,2$ and $a=3,4,5$. The action is invariant under worldsheet 
reparameterisations:

\beq
\delta_{\xi} X^{\alpha} = \xi^i\pl_i X^{\alpha}\ ,\quad
\delta_{\eta} X^a = \eta^i\pl_i X^a\label{wsrep}\ ,\qquad i=0,1\ .
\eeq

Note that, due to the absence of a worldsheet metric, there is no need to 
identify the vector fields $\xi$ and $\eta$. The equations of motion are:

\beq
\epsilon_{\alpha\beta\gamma} {\dot X}^{\beta}  X^{\prime\, \gamma} =0
\ ,\quad
\epsilon_{abc}  {\dot X}^b  X^{\prime\, c} =0\ .\label{fe}
\eeq

This means that the embedding of the worldsheet in the
0,1,2 directions is a one--dimensional submanifold of that
space. And similarly the embedding in the 3,4,5 directions is also
one--dimensional.

A special feature of our system is that both the equations of motion and
the gauge transformations are first order in derivatives. 
Thus, before we start the canonical analysis, it is instructive to 
first  analyse the solutions to the equations of motion. We assume that
the one--dimensional embedding of the worldsheet in the 0,1,2
directions is timelike. This means that we can fix the following static 
gauge for the $\xi$ symmetry:

\beq X^0 = \tau\ . \eeq

The field equation then implies that

\beq
X^{\prime\,1,2}=0\ .
\eeq

There are two inequivalent sectors of the theory which differ by how the 
one--dimensional embedding in the remaining 3,4,5 directions takes place.

\begin{enumerate}

\item[i)] The string sector, for which the gauge fixed $X^a$ field
 equation is

\beq \dot{X}^a=0\ . \label{david} \eeq

\item[ii)] The particle sector, for which the gauge fixed $X^a$ field
equation
is

\beq X^{\prime\,a}=0 \ .\eeq

\end{enumerate}

Hence, in the case of (i) the embedded worldsheet is a two--dimensional
surface which is unconstrained in the $X^{1,2}$ directions and fixed in the 
$X^{3,4,5}$ directions. In the case of (ii) the worldsheet is contracted 
into a one--dimensional worldline which is unconstrained in all five spatial 
directions. (By unconstrained we mean that the coordinate is pure
gauge; this will become apparent after the analysis described below.)
In the case of strings ending on D--branes the analogous field equations 
are much simpler, namely: $\dot{X}^i=0,\  i=1, \cdots ,p$. This leads 
to the notion 
of a zero--brane in the D--brane. Here however we find a membrane in the
five--brane due to the additional Dirichlet conditions (\ref{david}).
The boundary string is positioned in the space transverse to this membrane.

Let us continue by analysing the phase space dynamics of the three Euclidean 
coordinates $\vec{X}=(X^3,X^4,X^5)$. The canonical momenta are given by

\beq
\Pi_{a}(\sigma):= {\delta S \over \delta \dot{X}^{a}(\sigma)}
= -{h\over 3}\epsilon_{abc}X^b X^{\prime}{}^{c}\ .
\eeq

The non--trivial canonical Poisson brackets are:

\beq
\{X^a(\sigma),\Pi_b(\sigma')\}:=\delta^a_b\,\delta(\sigma-\sigma')\ .
\eeq

Since the Lagrangian is first order in time derivatives there is one primary 
constraint $\phi_a(\sigma)$ for each canonical momentum, given by

\beq
\phi_a := \Pi_a + {h \over 3} \epsilon_{abc}
X^b X^{\prime}{}^c  \approx 0\ .
\eeq

The canonical Hamiltonian $H := \vec{\Pi} \cdot \dot{\vec{X}}-L$ vanishes. 
Instead, one introduces a generalised Hamiltonian $ H_{{\rm gen}} := \int 
d\sigma \, \lambda^a(\sigma)\phi_a(\sigma)$ where $\lambda^a(\sigma)$ are 
three Lagrange multipliers. To proceed with the canonical analysis we study 
the consistency conditions

\beq
\dot{\phi}_a(\sigma) := \{ H_{\rm gen} , \phi_a(\sigma) \} =
\lambda^b(\sigma) M_{ba}(\sigma)\approx 0\, ,
\label{weak}
\eeq

where

\beq
\{\phi_a(\sigma) , \phi_b (\sigma') \}= M_{ab}(\sigma)
\delta(\sigma-\sigma') \ , \quad
M_{ab} = h\, \epsilon_{abc} X^{\prime}{}^c\ .
\eeq

Clearly, (\ref{weak}) imposes no further phase space constraints. It simply 
sets to zero the Lagrange multipliers in the directions where the matrix 
$M_{ab}$ is non--degenerate. These directions correspond to second class 
constraints. In a direction where $M_{ab}$ is degenerate the Lagrange 
multiplier remains undetermined and such a direction corresponds to a first 
class constraint. Hence, if $|\vec{X}^{\prime}| \equiv 0$ then there are 
three first class constraints and the phase space is trivial.

On the other hand, provided that $|\vec{X}^{\prime}| \neq 0$ the matrix 
$M_{ab}$ has only one zero eigenvector, given by $\lambda^a(\sigma)= 
\lambda(\sigma) X^{\prime}{}^{a}$. The matrix $M_{ab}$ is therefore 
non--degenerate in the two--dimensional subspace orthogonal to 
$\vec{X}^{\prime}$. Thus one introduces a projection onto this subspace
as follows ($I=1,2$):

\bea
P_I{}^a(\sigma)P_J{}^b(\sigma)\delta_{ab}&=&\delta_{IJ}\ ,\\
\delta^{IJ}P_I{}^a(\sigma)P_J{}^b(\sigma)&=&\delta^{ab}-{X^{\prime\, a}
X^{\prime\, b}\over
|\vec{X}'|^2} \ ,\\
\epsilon^{IJ}P_I{}^a(\sigma)P_J{}^b(\sigma)&=&
{\epsilon^{abc}X^{\prime\, c}\over |\vec{X}'|}\ .
\eea

The three constraints $\phi_a$ now split into the two second class 
constraints

\beq
\chi_I := P_I{}^a \phi_a\ ,
\eeq

with the now non--degenerate matrix

\beq
\{ \chi_I(\sigma) ,\chi_J(\sigma') \} := M_{IJ}(\sigma)
\delta(\sigma-\sigma') \ ,\qquad
M_{IJ} = P_I{}^a P_J{}^b M_{ab}\ ,
\label{mij}
\eeq

and one first class constraint

\beq
\phi := X^{\prime}{}^a\phi_a \equiv
X^{\prime\, a}\Pi_a\ ,
\eeq

which acts as the generator of $\sigma$--reparameterisations. The resulting 
generalised Hamiltonian,

\beq
H_{\rm gen} = \int d\sigma\, \lambda(\sigma) \phi(\sigma)\ ,
\eeq

leads to the canonical field equations

\beq
\dot{X}^a(\sigma) = \lambda(\sigma) X^{\prime}{}^a(\sigma)\ ,
\label{cfe}
\eeq

which are indeed consistent with the Lagrangian field equations (\ref{fe}) 
when $|\vec{X}^{\prime}| \neq 0$. We remark that the first class nature of the
constraint $\phi$ means that the gauge invariant dynamics is trivially 
realised (i.e.~ $\dot{V}=0$) on reparameterisation invariant functionals
$V[\vec{X}]$ satisfying

\beq
\int d\sigma\, \eta(\sigma) X'^a(\sigma){\delta V\over \delta X^a(\sigma)}
\approx 0\ .
\label{vrep}
\eeq

The analysis of the $X^{\alpha}$ coordinates is similar, but we require this
system to admit the solution $X^0=\tau$.  This implies, from the equations of
motion, that we must consider the trivial sector $X^{\prime\,\alpha}\equiv 0$.
(In this sector $X^0=\tau$ is a gauge choice for  the constraint
$\phi_0\approx0$.) 

To summarise, the two
admissible sectors of the theory given above are: (i) $X^{\prime \alpha}=0$ and
$|\vec{X}^{\prime}|=0$; and (ii) $X^{\prime \alpha}=0$ and
$|\vec{X}^{\prime}|\neq 0$.

Let us continue by deriving the symplectic structure of the phase
space. In the 0,1,2 directions there are no second class
constraints and we find vanishing brackets between the $X^{0,1,2}$ coordinates.
We next turn to the $X^{3,4,5}$ coordinates. In the string sector,
the presence of
the second class constraints leads us to define a Dirac bracket as follows:

\beq
\label{db}
[A,B]^{D} := \{ A , B \} - \int d\sigma \,
\{ A , \chi_I(\sigma) \} (M^{-1})^{IJ}(\sigma) \{ \chi_J(\sigma) , B \}\ .
\eeq

Here $A$ and $B$ are general phase space variables and
$(M^{-1})^{IJ}$ is the ordinary matrix inverse of the matrix $M_{IJ}$ 
defined in (\ref{mij}). The basic Dirac brackets between the coordinates 
$X^a$ and the first class constraint $\phi$ are:

\beq
[X^a(\sigma),X^b(\sigma')]^D =-{1\over h} {\epsilon^{abc}
X^{\prime}{}^c(\sigma)\over |\vec{X}^{\prime}(\sigma)|^2}
\delta(\sigma-\sigma')\ , \label{xx}
\eeq
\beq
[X^a(\sigma),\phi(\sigma')]^D = X^{\prime\, a}(\sigma)\delta(\sigma-\sigma')
\label{xphi}
\eeq
\beq
[\phi(\sigma),\phi(\sigma')]^D = 2\phi(\sigma)\delta'(\sigma-\sigma')+
\phi'(\sigma)\delta(\sigma-\sigma')\ .
\label{vir}
\eeq

From the point of view of the five--brane the fields $X^a$ are worldvolume
coordinates. The Dirac bracket (\ref{xx}) therefore describes a noncommutative
loop space on the five--brane.

Finally, we discuss three special topics: dimensional reduction, the 
canonical analysis in the infinite momentum frame and the Moyal
quantisation in case of a compact direction.
\bigskip

\subsection*{ Double Dimensional Reduction}

We wrap the five--brane and the membrane around a spacetime circle of radius 
$R$. In the limit of small $R$ the five--brane becomes a D4--brane and the 
wrapped membrane becomes a fundamental string. Thus one recovers the case of 
a string ending on a D4--brane in the background of constant Neveu--Schwarz
two--form potential. Explicitly, taking winding numbers into account, we find
for a five--brane winding $M$ times in spacetime and a string winding $N$ 
times in the five--brane

\beq
X^{5}(\sigma)=NR\sigma\, , \qquad
{\cal H}_{\mu \nu 5}={M\over R}{\cal F}_{\mu\nu}\ ,
\label{x5}
\eeq

where the index $\mu$ now refers to the D4--brane worldvolume and ${\cal F}$ 
is normalised such that it obeys $d{\cal F}=H^{NS}$ where $H^{NS}$ is the 
$D=10$ Neveu--Schwarz three--form field strength. The membrane winds $MN$ times in
spacetime and the membrane action (\ref{ac}) reduces to $MN S_{F1}$, where 
$S_{F1}$ is the $D=10$ string action (\ref{stringa}). For illustrative 
purposes, let us choose ${\cal H}_{345}=h$ and ${\cal H}_{012}=- 
h/\sqrt{1+\ell_p^6 h^2}$. For this choice we find a rank two magnetic 
field\footnote{This result can be generalized to a rank $4$ magnetic field 
by either reducing over a skew spacelike direction in the $0-5$ plane or 
boosting the constant ${\cal H}$ background in the $0-5$ direction turning 
on ${\cal H}_{034}$ and ${\cal H}_{125}$ components, and then reduce over 
the fifth direction.}

\beq
{\cal F}_{34}={hR\over M}\ ,
\label{f34}
\eeq

leading to a noncommutative D--brane worldvolume with

\beq
[X^3,X^4]={1\over MN} ({\cal F}^{-1})^{34}\ .
\label{th34}
\eeq

Alternatively, the double dimensional reduction can be performed directly at
the level of the algebra (\ref{xx})-(\ref{vir}) by fixing the gauge 
$X^{5}(\sigma)= NR\,\sigma$. The gauge fixed Dirac bracket 
$[\cdot,\cdot]^{D'}$ reads

\beq
[X^3(\sigma),X^4(\sigma')]^{D'}=-{1\over h
NR}\delta(\sigma-\sigma')\ .
\label{xxgf}
\eeq

This bracket reduces to (\ref{th34}) provided we make use of (\ref{f34}) and 
identify $X^{3,4}$ with the zero--modes $ X^{3,4}= \int d\sigma 
X^{3,4}(\sigma)$ of the doubly reduced string.

\bigskip

{\subsection*{ Canonical Analysis in the infinite momentum frame}}

In the infinite momentum frame discussed in Appendix C the field strength
${\cal H}$ is degenerate along a null direction:

\beq
{\cal H}_{-\mu\nu}=0\ .
\label{deg}
\eeq

The remaining components of the field strength are given by (see
Eq.~(\ref{llfs}) in Appendix C):

\beq
{\cal H}_{+pq}=F^-_{pq}=-{1\over 2}\epsilon_{pqrs}F^-_{rs}\ ,\quad {\cal
H}_{pqr}=0\ ,\quad p=1,2,3,4\ .
\eeq

Thus the lightcone coordinate $X^-$ is decoupled from the remaining fields
$X^+$ and $X^p$ in the string action (\ref{action0}). The action has the
reparameterisation invariance:

\beq
\delta_{\eta}X^+=\eta^i\pl_i X^+\ , \quad \delta_{\eta}X^p=\eta^i\pl_i X^p\ .
\eeq

In the non--trivial sector the embedding of the worldsheet is
a two--dimensional surface and the gauge fixed solutions to the field
equations  can be taken to be

\beq
X^-=\sigma^-\ ,\quad \pl_-X^+=\pl_-X^p=0\ ,
\eeq

where $\sigma^{\pm}={1\over \sqrt{2}}(\tau\pm\sigma)$. To give a phase space
formulation, we introduce canonical momenta with respect to two--dimensional
lightcone velocities as follows:

\beq
\Pi_{\mu}(\sigma^+):={\delta S\over \delta \pl_-X^{\mu}(\sigma^+)}\ ,
\quad \mu = +,p\ .
\eeq

The first order nature of the Lagrangian implies that there are five primary
constraints

\beq
\phi_{\mu}:=\Pi_{\mu}+{1\over 3} {\cal H}_{\mu\nu\rho}X^{\nu}\pl_+ X^{\rho}
\approx 0 \ , \quad \mu = +,p\ .
\eeq

Since $F^-_{pq}$ has rank four, it follows that the bracket
$\{\phi_{\mu},\phi_{\nu}\}$ has rank four and zero direction given by $\pl_+
X^{\mu}$. Hence there is one first class constraint, that we can choose to
be

\beq
\phi:=\pl_+ X^{\mu}\phi_{\mu}\equiv \pl_+ X^{\mu}\Pi_{\mu}\ ,\quad \mu=+,p\ ,
\eeq

and four second class constraints $\chi_I$, $I=1,2,3,4$. Provided that $\pl_+
X^{+}\neq 0$, we can choose these to be

\beq
\chi_I=\delta_I^p\phi_p\ .
\eeq

The basic Dirac brackets are found to be

\bea
[X^p(\sigma^+_1),X^q(\sigma^+_2)]^D&=&{[(F^-)^{-1}]^{pq}\over
\pl_+ X^{+}}\delta(\sigma^+_1-\sigma^+_2)\ ,
\cr&&\cr
[X^+(\sigma^+_1),X^{\mu}(\sigma^+_2)]^D&=&0\ ,\quad \mu=+,p\ ,
\eea

and the analogs of (\ref{xphi}) and (\ref{vir}).

In a background with a compact light--like direction of radius $R$, the
gauge choice $X^+=NR\sigma^+$ leads to a local, field independent bracket
analogous to (\ref{xxgf}):

\beq
[X^p(\sigma^+_1),X^q(\sigma^+_2)]^D = {[(F^-)^{-1}]^{pq}\over
NR}\delta(\sigma^+_1-\sigma^+_2)\ .\label{extra}
\eeq

A double dimensional reduction
along the compact direction gives the Dirac bracket for the rank $4$
noncommutative D4--brane with Born-Infeld field strength ${\cal
F}_{\mu\nu}={R\over M}\delta_{\mu}^p\delta_{\nu}^q F^-_{pq}$ where $M$ is
the number of times the five--brane winds around the compact direction.
\bigskip

\subsection*{ Moyal quantisation}

Usually after deriving the noncommutative structure one determines the star 
product. This cannot be done in a direct way for the bracket (\ref{xx}). We 
will discuss a proposal for how this might be done in the following 
discussion section.
However, in the special case where the five--brane has a compact spacelike
direction, the gauge fixed Dirac brackets (\ref{xxgf}) are straightforward to
quantise by introducing the following Moyal product $(I,J=3,4)$:

\beq
F(X)\star G(X) =
\exp\left[-{1\over 2hNR}\int d\sigma \epsilon^{IJ}{\delta\over\delta
X^I(\sigma)}
{\delta\over\delta Y^J(\sigma)} \right] F(X)G(Y)\vert_{X=Y}\ ,
\label{moyal}
\eeq

where $F$ and $G$ are functionals of $X^{3,4}$.

Similarly, in 
a background with a compact light--like direction 
of radius $R$, the Moyal quantised star--product, based upon
the brackets (\ref{extra}), is given by ($p=1,2,3,4)$:

\beq
F[X]\star G[X]=\exp\left[-{1\over 2NR}\int d\sigma^+\,[(F^-)^{-1}]^{pq}
{\delta\over \delta X^p(\sigma^+)} {\delta\over \delta Y^q(\sigma^+)}
\right]
F[X] G[Y]|_{X=Y}\ ,
\label{moya}
\eeq

where $F$ and $G$ are functionals of $X^p$.

\section{Discussions}

In this paper we have studied the effect of having a constant $C$--field on 
the M--theory five--brane using an open membrane probe. We proposed a 
specific decoupling limit in which the open membrane metric remains 
non--degenerate. An unconventional feature of this decoupling limit, not 
encountered in the case of a string probing a D--brane, is that the open 
membrane action scales with an overall factor, as was discussed in Section 
IV.

We hope that our work provides the motivation and initial steps to study 
noncommutative loop algebras. A natural question to ask in this context is 
whether we can define a star product like in the case of D--branes.
The utility of the star product is that to incorporate the effects of the 
noncommutativity of the space on fields one simply replaces ordinary 
products with star products to give the deformed theory.

In the case of a point particle moving in a finite dimensional Poisson 
manifold $M$ with Poisson structure $\Omega^{ij}$, one may in principle 
always find local canonical coordinates in which $\Omega^{ij}$ is a constant 
matrix and the star product then becomes simple. The reparameterisation 
independent definition of the star product was first given by Kontsevich 
\cite{Kontsevich:1997vb} and has recently been cast into a more physical 
context by the work of \cite{Cattaneo}. The latter work makes use of a path 
integral representation of the star product.

There is a natural extension of the definition of star product given in 
\cite{Cattaneo} that applies to the case of the open membrane/five--brane 
system. A physically intuitive definition of the star product of two loop 
functionals $F[X]$ and $G[X]$ is given by the following path integral 
expression

\beq
(F\star G)[X(\sigma)] = \int\limits_{\lim\limits_{\tau\rightarrow \pm
\infty}X(\tau,\sigma)=X(\sigma)} {\cal D}X
~F[X(\tau=1,\sigma)] ~G[X(\tau=0,\sigma)] ~\exp{{i\over \hbar} S[X]}\ ,
\label{fg}
\eeq

with the action S  given by (\ref{wza}). In case the five--brane winds 
around a compact direction in the background we expect (\ref{fg}) to 
reproduce the Moyal products given in (\ref{moyal}) and (\ref{moya}).

It would be interesting to see whether the proposed path integral 
representation of the deformed five--brane geometry will facilitate the 
construction of a loop space analog of the star product of ordinary fields. 

It will be desirable to represent the loop algebra also on ordinary fields 
such as those of the massless $(2,0)$ tensor multiplet on the five--brane. 
One possible direction for such constructions may be to introduce so called 
loop space covariant derivatives \cite{Bergshoeff:1991ei}, which naturally 
incorporates the two--form potential in the tensor multiplet.

The main motivation for introducing a noncommutative loop algebra in this 
work was provided by a study of the M--theory five--brane. However, there 
are also other reasons to study the same system. For instance, the action 
(\ref{action0}) is formally the same as the action for a string interacting 
with a linear background Neveu--Schwarz two--form. Here we identify $B^{\rm
NS}_{\mu \nu}= { 1 \over 3} {\cal H}_{\mu \nu \rho} X^{\rho}$ such that the
Neveu--Schwarz field strength, $H^{\rm NS}= d B^{\rm NS}$ is a constant.
However, despite  the resemblance of the terms in the action, the conditions
on the background  fields differ. In the previous case ${\cal H}$ must obey
the field equations  of the five--brane, in the case of the string with a
background Neveu--Schwarz field strength, the background fields must obey the
Einstein's equations for the appropriate supergravity. One can construct a
solution  with constant Neveu--Schwarz field strength, but the space is
necessarily curved and one must again take care when discussing limits.

It is interesting that the action (\ref{action0}) also arises from the
dimensional reduction of the  higher order Chern--Simons term that appears in
$11$--dimensional  supergravity $S=\int_{M^{11}} C \wedge H \wedge H$. More
precisely, take  $M^{11}=M^2\times M^9$ and then reduce on $M^9$ as follows: 
Let $C=X^I  \gamma_I$ where $\gamma_I \in H^{3}(M^9) \, , \, I=1 \cdots
b^3(M^9)$ and  $X^I$ are scalars on $M^2$, one then identifies $H$ as the
triple  intersection form on $M^9$, $H_{IJK}= \int_{M^9} \gamma_I \wedge 
\gamma_J\wedge \gamma_K$. For this case the only constraints on the 
background field $H$ is that it is a triple intersection form. The kinetic 
terms of the supergravity will introduce period matrices of $\{\gamma_I\}$ 
that will play the role of a metric for $X^I$. In $5$--dimensional 
supergravity there is an analogous treatment of the $A_1 \wedge F_2\wedge 
F_2$ term. It would be interesting to see whether (compactified) 
supergravity theories allow limits in which only the Wess--Zumino terms
survive.

It is worth remarking that in the decoupling limit we defined in this work, 
it appears possible to take seriously a quantisation program for the 
membrane. This is because one has decoupled the bulk modes and so one is 
left with only the boundary string in the five--brane. It would be 
interesting to see how far one
can carry out the quantisation of such a system. In doing this one should 
use the full supersymmetric branes.

Finally, it is known that for the case of the D--brane the noncommutative 
U(1) theory is non--abelian. It is our hope that the development of the 
noncommutative M5--brane might shed light on the as yet unknown non--abelian 
structure of the five--brane worldvolume theory.
\bigskip

{\bf Note Added:} After this paper appeared, we received the work of
\cite{Kawamoto:2000zt} in which related ideas are discussed from a different
point of view.

\section*{Acknowledgements}

We would like to thank Martin Cederwall, Chong-Sun Chu, 
Christiaan Hofman, Whee Ky Ma and Ergin Sezgin for discussions. 
D.B. would like to thank the organisers and participants of the Leiden 
conference on Noncommutative Geometry. This work is supported by the 
European Commission TMR program ERBFMRX-CT96-0045, in which E.B., D.B.~and
J.P.v.d.S.~are associated to the University of Utrecht. The work of 
J.P.v.d.S.~and P.S.~is part of the research program of the ``Stichting voor
Fundamenteel Onderzoek der Materie'' (FOM).

\appendix

\section{The Wess--Zumino term of the membrane }

In this appendix we derive the form of the Wess--Zumino term of the membrane 
as given in (\ref{form}).

Assume that $f_5^* F=0$ and $\cal H$ is constant. Then we can write $f_5^* C 
= dc$ and ${\cal H}= d(b+c)$, at least locally. The two Wess--Zumino terms 
in (\ref{ac}) can therefore be written as a Wess--Zumino term for the bulk 
modes perpendicular to the five--brane plus a string Wess--Zumino term as 
follows:

\beq
\int_{M^3}f_2^* \tilde{C}+\int_{\partial M^3} f_1^*(b+c)\ ,
\eeq

where $\tilde{C}$ obeys $f_5^*\tilde{C}=0$. To see this, one writes 
$C=\tilde{C}+dC_2$ where $f_5^* C_2=c$. One then applies Stokes' theorem to 
$\int_{M^3}C$ and uses the identity

\beq
f_2^\star C_2 |_{\partial M^3} = f_1^\star f_5^\star C_2 = f_1^\star c\, .
\eeq

Finally, one uses the fact that, since ${\cal H}$ is constant, we have
the identity

\beq
(b+c)_{\mu\nu} = {1\over 3}{\cal H}_{\mu\nu\rho}X^\rho\, .
\eeq

Combining everything then leads to the form (\ref{form}) of the
Wess--Zumino term for the string in the five--brane.

\section{The $SO(5,1)/SO(2,1)\times SO(3)$ parameterisation of ${\cal 
H}_{\mu\nu\rho}$}

In this appendix we parameterise a generic solution to the non--linear 
self--duality condition (\ref{nlsd}) using a real scale factor and a coset 
element in $SO(5,1)/SO(2,1)\times SO(3)$. The special solutions 
corresponding to an infinite momentum frame are described in the next 
appendix.

The non--linear self--duality condition (\ref{nlsd}) is equivalent to a 
linear self--duality condition \cite{Howe:1997mx}

\begin{eqnarray}
{\cal H}_{\mu\nu\rho} &=&4(m^{-1})_{\mu}{}^\lambda h_{\nu\rho\lambda}
\ ,\quad
h_{\mu\nu\rho}={\sqrt{-g}\over 6}\epsilon_{\mu\nu\rho\sigma\lambda\tau}
h^{\sigma\lambda\tau}\, ,\nonumber\\
m_{\mu\nu} &=&g_{\mu\nu}-2\ell_p^6\, k_{\mu\nu}
\ ,\hskip 1.3truecm  k_{\mu\nu}=h_{\mu\lambda\rho}h_\nu{}^{\lambda\rho}\ .
\label{sd}
\end{eqnarray}

It can
be shown that the matrix $k_{\mu\nu}$ is traceless and that its square is 
proportional to $\eta_{\mu\nu}$. Hence $k_{\mu\nu}$ can be written as

\beq
k_{\mu\nu}= \lambda_+ v^+_{\mu} v^+_{\nu}+\lambda_- v^-_{\mu}v^-_{\nu}
+\lambda\sum_{I=1}^2(-e^I_{\mu}e^I_{\nu}+e^{I+2}_{\mu}e^{I+2}_{\nu})\ ,
\eeq

where the three real parameters $\lambda_{\pm}$ and $\lambda$ obey 
$\lambda_+\lambda_-=\lambda^2$ and $(v^{\pm}_{\mu},e^p_{\mu})$, $p=1,2,3,4$, 
is an element of the coset $SO(5,1)/SO(1,1)\times SO(4)$ defined by

\beq
v^{\pm\,\mu}v^{\pm}_{\mu}=\eta^{\pm\pm}\ ,\quad
v^{\pm\,\mu}e^p_{\mu}=0\ ,
\quad e^{p\mu}e^q_{\mu}=\delta^{pq}\ ,
\label{42}
\eeq
\beq
\eta^{+-}=\eta^{-+}=-1\ ,\quad \eta^{++}=\eta^{--}=0\ ,\quad
\epsilon^{-+pqrs}=\epsilon_{+-pqrs}= \epsilon^{pqrs}\ .
\label{etapm}
\eeq

If $\lambda_{\pm}\neq 0$ (the cases $\lambda_+=\lambda=0$ and 
$\lambda_-=\lambda=0$ correspond to the infinitely boosted solutions 
discussed in the next appendix) we can use a boost to set 
$\lambda_+=\lambda_-=\lambda$ such that

\beq
k_{\mu\nu}=-\lambda \eta_{\alpha\beta}v_\mu^{\alpha}v_\nu^{\beta}
+\lambda\delta_{ab}u_\mu^{a}u_\nu^{b}\ ,
\eeq

where $(v^{\alpha}_{\mu},u^a_{\mu})$, $\alpha=0,1,2$, $a=3,4,5$, is an 
element of the coset $SO(5,1)/SO(2,1)\times SO(3)$ obeying (\ref{sol}). 
Keeping the coset element fixed, any three--form can be expanded in this 
basis as follows

\beq
\omega_{\mu\nu\rho}=A u^3_{\mu\nu\rho} + B^{a}_{\alpha}
(u^2v)_{a}^{\alpha}{}_{\mu\nu\rho}
+ C_{a}^{\alpha} (uv^2)^{a}_{\alpha}{}_{\mu\nu\rho}
+ D v^3_{\mu\nu\rho}\ ,
\label{gf}
\eeq

where the coefficients together make up twenty real components and we have 
defined the following basis elements for three--forms:

\bea
u^3_{\mu\nu\rho}&=&\epsilon_{abc}u^{a}_\mu u^{b}_\nu u^{c}_\rho
\ ,\quad (u^2v)_{a}^{\alpha}{}_{\mu\nu\rho}=
\epsilon_{abc}u^{b}_{[\mu}u^{c}_\nu v^{\alpha}_{\rho]}
\ ,\\
(uv^2)^{a}_{\alpha}{}_{\mu\nu\rho}&=&\epsilon_{\alpha\beta\gamma}
u^{a}_{[\mu}v^{\beta}_\nu v^{\gamma}_{\rho]}\ ,\quad
v^3_{\mu\nu\rho}=\epsilon_{\alpha\beta\gamma}v^{\alpha}_\mu v^{\beta}_\nu
v^{\gamma}_\rho\ .
\eea

For our choice of Lorentz basis, the Hodge $\star$ acts as follows

\beq
\star u^3 = v^3 \ ,\quad \star
(uv^2)^{a,\alpha}=(u^2v)^{a,\alpha}\ .
\eeq

Hence, we can write a general self--dual tensor in the form (\ref{gf}) using 
the ten coefficients $A=D$ and $B^{a,\alpha}=C^{a,\alpha}$. In particular, 
we can apply this expansion to $h_{\mu\nu\rho}$ itself, which yields the 
following expression for $k_{\mu\nu}$:

\bea
k_{\mu\nu}&=&2A^2(u^2_{\mu\nu}-v^2_{\mu\nu}) +
{2\over 9}\left[ \left( B^2\delta_{ab}-2
B^2_{ab}\right)
u^{a}_\mu u^{b}_\nu +
\left( B^2 \eta_{\alpha\beta} + 2 B^2_{\alpha\beta}\right)
v^{\alpha}_\mu v^{\beta}_\nu \right]  \nonumber \\
&&+{4\over 9} \epsilon_{abc} \epsilon_{\alpha\beta\gamma}
B^{a,\alpha} B^{b,\beta} u_{(\mu}^{c}v_{\nu)}^{\gamma}\ ,
\eea

where we have written $B^2_{\alpha\beta}=\delta^{ab} B_{a,\alpha} 
B_{b,\beta}$, $B^2_{ab}=\eta^{\alpha\beta}B_{a,\alpha} B_{b,\beta}$ and 
$B^2=\delta^{ab}B^2_{ab}=\eta^{\alpha\beta}B^2_{\alpha\beta}$. However, by 
assumption the matrix $k_{\mu\nu}$ must be diagonal in this basis. This 
implies that there must exist real numbers $X$ and $Y$ such that

\beq
B^2_{ab}=X\delta_{ab}\ ,\quad B^2_{\alpha\beta}=Y\eta_{\alpha\beta}\ .
\label{bxy}
\eeq

This implies that $B_{a,\alpha}$ has to vanish since it is an irreducible 
representation of $SO(3)\times SO(2,1)$. This can be verified explicitly by 
first taking the determinant of the equations in (\ref{bxy}) which shows 
that $X=Y<0$, and then deducing from $B^2_{1,1}=B^2_{2,2}=Y<0$ that 
$B_{a,1}=B_{a,2}=0$ and hence $X=Y=0$. It then follows from $B^2_{0,0}=0$ 
that also $B_{a,0}=0$. Substituting $B_{a,\alpha}=C_{a,\alpha}=0$ into 
Eq.~(\ref{gf}) gives

\beq
h_{\mu\nu\rho}=A \left(u^3_{\mu\nu\rho} + v^3_{\mu\nu\rho}\right)\ .
\eeq

From this is follows that the matrix $m_{\mu\nu}$ in Eq.~(\ref{sd}) is given
by

\beq
m_{\mu\nu} = (1-4\ell_p^6 A^2)u^2_{\mu\nu}
+ (1+4\ell_p^6 A^2) v^2_{\mu\nu}\, .
\eeq

After application of (\ref{sd}) one finally obtains the expressions 
(\ref{hsol}) and (\ref{openm}) for the non--linearly self--dual three--form 
and the open membrane metric provided one makes the following 
identification:

\beq h = {4A\over 1-4\ell_p^6\, A^2}\, .
\eeq

\section{The decoupling limit in the infinite momentum frame}

In this appendix we discuss a modified version of the decoupling limit of 
Section IV where a boost parameter is scaled such that it becomes infinite 
in the limit. This results in a decoupled string action of the form 
(\ref{action0}) where the constant background field strength is given by 
applying the infinite boost to the generic solution (\ref{hsol}). The 
infinitely boosted field strength is most naturally described in terms of an 
$SO(5,1)/SO(1,1)\times SO(4)$ coset element (\ref{42}) and a self--dual 
two--form in the four--dimensional Euclidean space perpendicular to the 
boost direction.

\subsection*{The $SO(5,1)/SO(1,1)\times SO(4)$ parameterisation of
${\cal H}_{\mu\nu\rho}$}

Given a coset element (\ref{42}) an arbitrary three--form $h_{\mu\nu\rho}$ 
can be parameterised in terms of two anti--symmetric matrices $F_{pq}$ and
$F'_{pq}$ and two vectors $G_p$ and $G'_p$ by giving its components with the 
respect to the frame as follows ($p=1,2,3,4)$:

\beq
h_{+pq}=F^-_{pq}+F^{\prime\,+}_{pq}\ ,\quad
h_{-pq}=F^+_{pq}+F^{\prime\,-}_{pq}\ ,
\eeq
\beq
h_{+-p}=G_p+G'_p\ ,\qquad h_{pqr}=\epsilon_{pqrs}(G_s-G'_s)\ ,
\eeq

where $F^{\pm}_{pq}={1\over 2}(F_{pq}\pm\epsilon_{pqrs}F_{rs})$ and 
$F^{\prime\pm}_{pq}={1\over 2}(F'_{pq}\pm\epsilon_{pqrs}F'_{rs})$. A 
self--dual three--form is obtained by setting $F'_{pq}=G'_p=0$. By an $SO(5,1)$
rotation we can fix a coset frame such that $G_p=0$. In this adapted frame 
the self--dual three--form then has the expansion

\beq
h_{\mu\nu\rho} = 3 v_{[\mu}^+e_\nu^p e_{\rho]}^q F_{pq}^-
+ 3 v_{[\mu}^-e_{\nu}^p e_{\rho]}^q F_{pq}^+
=3v^+_{[\mu}F^-_{\nu\rho]}+3v^-_{[\mu}F^+_{\nu\rho]}\ ,
\eeq

where $F^{\pm}_{\mu\nu} = e^p_{\mu}e^q_{\nu}F^{\pm}_{pq}$. The local 
$SO(1,1)\times SO(4)$ symmetry can be fixed by taking 
$v^{\pm}_{\mu}=(1,\pm\hat{n})$, where $\hat{n}$ is a unit vector in ${\bf
R}^5$, and choosing $e^p_{\mu}$ such that
$F_{pq}^{\pm}=f_{\pm}(i\sigma^2\oplus(\pm i\sigma^2))_{pq}$, where $f_{\pm}$
are two real parameters. This makes a total of ten independent degrees of 
freedom. From (\ref{sd}) and (\ref{om}) it follows that the non--linearly 
self--dual three--form and the open membrane metric are given in this
parameterisation by

\beq
{\cal H}_{\mu\nu\rho}={12\over 1-4\ell_p^{12}(F^+)^2(F^-)^2}\left[
v^+_{[\mu}F^-_{\nu\rho]}+v^-_{[\mu}F^+_{\nu\rho]}
-2\ell_p^6(F^-)^2v^+_{[\mu}F^+_{\nu\rho]}-2\ell_p^6(F^+)^2v^-_{[\mu}F^-_{\nu\rho]}\right]
\ ,
\eeq
\bea
G_{\mu\nu}={1\over \left(1-4\ell_p^{12}(F^+)^2(F^-)^2\right)^2}&&\left[
\left(1+4\ell_p^{12}(F^+)^2(F^-)^2\right)g_{\mu\nu}\right.\nonumber\\
&+&\left.4\ell_p^6 v^+_{\mu}v^+_{\nu}(F^-)^2
+4\ell_p^6 v^-_{\mu}v^-_{\nu}(F^+)^2-16\ell_p^6 
F^-_{\mu\rho}F^+_{\nu}{}^{\rho}
\right]\ .
\eea

where we have defined $(F^{\pm})^2=F^{\pm\mu\nu} F^{\pm}_{\mu\nu}=F^{\pm 
pq}F^{\pm}_{pq}$. Provided that $F^{\pm}_{\mu\nu}\neq0$ this 
parameterisation is equivalent to the parameterisation in (\ref{hsol}). An
infinite boost along the direction $\hat{n}$ is obtained by setting 
$F^+_{\mu\nu}=0$ (corresponding to the special case $\lambda_-=\lambda=0$ 
mentioned under Eq.~(\ref{etapm})) and results in the expressions

\bea
{\cal H}_{\mu\nu\rho}&=&12v^+_{[\mu}F^-_{\nu\rho]}\ ,\label{imf}\\
G_{\mu\nu}&=&g_{\mu\nu}+4\ell_p^6 v^+_{\mu}v^+_{\nu}(F^-)^2\ .
\label{llf}
\eea

We notice that (\ref{imf}) is actually linearly self--dual and the second 
term in (\ref{llf}) drops out of the non--linear condition (\ref{nlsd}). 
More explicitly, in the infinite momentum frame the components of ${\cal H}$ 
are

\beq
{\cal H}_{+pq}=4F^-_{pq}\ ,\quad {\cal H}_{-pq}=0\ ,\quad
{\cal H}_{pqr}={\cal H}_{+-p}=0\ .
\label{llfs}
\eeq

\subsection*{The Infinite Momentum Decoupling Limit}

In order to define a decoupling limit leading to (\ref{action0}) with the 
field strength given by (\ref{imf}), we first boost (\ref{hsol}) along the 
$5$ direction:

\beq
\left(\begin{array}{l} v^0\\u^5\end{array}\right)=
\left(\begin{array}{cc} \sqrt{1+a^2}&a\\a&\sqrt{1+a^2}\end{array}\right)
\left(\begin{array}{l} \tilde{v}^0\\\tilde{u}^5\end{array}\right)\ ,
\eeq

and then keep $\tilde{v}^0$ and $\tilde{u}^5$ fixed, while scaling according
to (\ref{nr}) and

\beq
a\sim \epsilon^{-\gamma}a\ ,\quad \gamma>0\ .
\eeq

The Wess--Zumino term now has weight $-\gamma-\Delta$, such that in the
rescaled action  (where the Wess--Zumino term is fixed) the kinetic energy
of the parallel modes has weight $-2+\delta-1+\gamma+\Delta>0$ and the kinetic
energy of the perpendicular modes has weight $-2+\gamma+\Delta<0$.
The open membrane metric becomes

\bea
G_{\mu\nu}&=&{\bigl (1+\sqrt{1+h^2\ell_p^6}\bigr )^2\over 4}
\left[\delta_{ab}u^{a}_{\mu}u^{b}_{\nu}
+{1 \over 1+h^2\ell_p^6}\eta_{\alpha\beta}v^{\alpha}_{\mu}v^{\beta}_{\nu}
\right.
\\
&+&\left.{a^2h^2\ell_p^6\over 1+h^2\ell_p^6}\left(
v^0_{\mu}v^0_{\nu}+2\sqrt{1+{1\over a^2}}v^0_{(\mu}u^5_{\nu)} +
u^5_{\mu}u^5_{\nu}\right)\right]\ ,
\eea

where we have dropped the tildes on $v^0$ and $u^5$. Requiring the leading 
components of this matrix to form a rank six matrix, we see that 
$a^2h^2\ell_p^6$ has to be finite, i.e.~ $\gamma+\Delta+{3\over
2}(\delta-1)\leq 3$. Thus the conditions on $\gamma+\Delta$ are the same as
the conditions on $\Delta$ given in (\ref{limsum}). Since
$h\ell_p^3\rightarrow 0$, the result of the limit is the decoupled boundary
Wess--Zumino term (\ref{action0}) with background field strength now given by
the infinitely boosted solution (\ref{imf}) provided we identify

\beq
v^{+}_{\mu}={1\over \sqrt{2}}(v^0_{\mu}+ u^5_{\mu})\ ,\quad
F^{-}_{\mu\nu}={ah\over \sqrt{2}}(- v^1_{[\mu}v^2_{\nu]}+ u^3_{[\mu}u^4_{\nu]})\ .
\eeq

\end{document}